\documentclass[pre,twocolumn,aps,floatfix,longbibliography,nofootinbib]{revtex4-2}
\usepackage{hyperref}
\usepackage{pslatex,graphicx,dcolumn,bm,natbib,amssymb,amsmath,color,mathtools}

\usepackage[utf8]{inputenc}
\usepackage[T1]{fontenc}
\usepackage{soul}
\usepackage{float}
\usepackage{comment}
\usepackage{ulem}
\usepackage{times}
\newcommand{\be}{\begin{equation}}
\newcommand{\ee}{\end{equation}}
\DeclareTextFontCommand{\cf}{\fontfamily{phv}\selectfont}
\newcommand{\state}[1]{\textbf{\cf{#1}}}

\makeatletter
\def\frontmatter@thefootnote{%
 \altaffilletter@sw{\@fnsymbol}{\@fnsymbol}{\csname c@\@mpfn\endcsname}%
}%
\makeatother

\begin{document} 
\title{Origin of yield stress and mechanical plasticity in model biological tissues}

\author{Anh Q Nguyen}
\author{Junxiang Huang}
\author{Dapeng Bi}
\affiliation{Department of Physics, Northeastern University, Boston, MA 02115, USA}
\affiliation{Center for Theoretical Biological Physics, Northeastern University, Boston, Massachusetts 02215, USA}

\begin{abstract}
    During development and under normal physiological conditions, biological tissues are continuously subjected to substantial mechanical stresses. In response to large deformations cells in a tissue must undergo multicellular rearrangements in order to maintain integrity and robustness. However, how these events are connected in time and space remains unknown. Here, using computational  and theoretical modeling, we studied the mechanical plasticity of epithelial monolayers under large deformations. Our results demonstrate that the jamming-unjamming (solid-fluid) transition in tissues can vary significantly depending on the degree of deformation, implying that tissues are highly unconventional materials. Using analytical modeling, we elucidate the origins of this behavior. We also demonstrate how a tissue accommodates large deformations through a collective series of rearrangements, which behave similarly to avalanches in non-living materials. We find that these ‘tissue avalanches’ are governed by stress redistribution and the spatial distribution of vulnerable spots. Finally, we propose a simple and experimentally accessible framework  to predict avalanches and infer tissue mechanical stress based on static images.
\end{abstract}

\maketitle

\section{Introduction}
During morphogenesis and under normal physiological conditions, biological tissues continuously experience substantial mechanical stresses  ~\cite{bonfanti_review_fracture_2022}. Research efforts to understand the remarkable deformability of epithelial tissues employ both experimental and simulation approaches. Experimentally, studies focus on the tissue’s responses to external stresses ~\cite{Harris_2012,monolayer_stretching,shear_experiment_Sadeghipour_2018,Bashirzadeh_2018,Cai_2022}, where a stress-driven unjamming transition has been noted  ~\cite{Cai_2022}. On the simulation front, the cellular Potts and Vertex-based models are utilized to probe tissue rheology  ~\cite{villemot2021quasistatic,merzouki2016_nonlinear_res}, uncovering nonlinear elasticity and rheological properties  ~\cite{merzouki2016_nonlinear_res,huang2022shear}. However, with few exceptions  ~\cite{hertaeg2024discontinuous}, research have predominantly focused on the shear startup regime. This leaves a gap in our understanding of tissue behavior under steady shear and the mechanisms underpinning yield-stress behavior in tissues.
Beyond the yield stress, materials typically flow through plastic rearrangements. Similarly, within tissues, mechanical plasticity occurs through cellular rearrangements, enabling the maintenance of integrity and robustness. While there is extensive literature on how individual cells rearrange with their neighbors  ~\cite{lienkamp2012vertebrate,10.7554/eLife.07090,HELLER2014617,CURRAN2017480}, significant gaps remain in understanding how these localized events connect over time and space. Moreover, a major challenge lies in elucidating how these collective interactions lead to mechanical responses at the tissue level. In the context of material plasticity, avalanche-like behavior, prevalent in phenomena ranging from earthquakes to ferromagnets, involves small perturbations triggering significant collective responses  ~\cite{ruina1983slip}. Systems exhibiting these instabilities display self-organized criticality  ~\cite{bak1987self_critical} and power law scaling in their observables, indicating the universality class of the process. {Proliferation-driven avalanche-like behavior has recently been studied using numerical simulation of the Drosophila eye disc, suggesting that avalanches provide a macroscopic mechanism for epithelial tissues to alleviate accumulated proliferative stress  ~\cite{courcoubetis_avalanches_by_proliferation}. Additionally, there is suggestive evidence of motility-induced avalanches in the epithelial tissue of Trichoplax adhaerens  ~\cite{Prakash_2021}, where dynamic forces from the organism’s motility trigger localized microfractures that cascade into larger fractures, a hallmark of avalanche dynamics.} Moreover, shear-induced avalanches have been documented in vertex-based models {first in   ~\cite{Popovic_2021}} and also   ~\cite{huang2022shear,hertaeg2024discontinuous}, yet a detailed examination of these avalanches’ growth and {evolution} is still lacking. 

In this work, we investigate tissue mechanical plasticity using the Voronoi-based Vertex model under quasi-static shear. Our results demonstrate that the solid-fluid transition point—also referred to as the jamming-unjamming transition in recent literature ~\cite{bi_nphys_2015,Bi_PRX_2016,irradiation_jamming}—does not occur at a singular point but varies depending on the degree of shear deformation the tissue undergoes. Furthermore, challenging traditional definitions, we discover states where tissues possess yield-stress properties but lack a conventional shear modulus. These states exist in a solid-fluid coexistence phase near the jamming-unjamming transition, which we explore through a modified version of the Soft Glassy Rheology model to elucidate the origins of these complex states. {The coexistence of fluid and solid phases suggests that even traditionally fluid-like tissues can accumulate stress in response to deformation, highlighting the need for a parameter-free, model-independent tissue stress inference method. Here, we propose a easily implementable metric to estimate tissue stress from static snapshots of tissue configuration without requiring any prior knowledge of the system.} Additionally, our research not only clarifies how tissue manages large deformations through multicellular rearrangements akin to avalanches observed in non-living materials but also connects these phenomena to the tissue-level mechanical responses discussed earlier. These “tissue avalanches” are driven by stress redistribution and the spatial distribution of {soft} spots ~\cite{falk2011deformation,Manning_Liu_2011,Patinet_Falk_2016,yang2024multicell}, elements that echo the earlier discussions on mechanical responses and rheological properties. By quantifying the spatiotemporal correlations within these rearrangements, we  propose a new methodological framework capable of predicting collective rearrangements.

\section*{Results}
 \subsection{The confluent jamming transition is not unique}

To investigate the mechanical behavior of dense epithelial tissues under substantial deformation, we employed a Voronoi-based Vertex model  ~\cite{Bi_PRX_2016,huang2022shear}. The cell centers $\{\bm r_i\}$ and their geometric configurations are derived from Voronoi tessellation. The biomechanical interactions are captured through a dimensionless mechanical energy functional  ~\cite{Staple_EPJE_2010} expressed as:
$
\varepsilon = \sum_{i=1}^N \left[\kappa_A (a_i - 1)^2 + (p_i - p_0)^2\right],$ where {$a_i$ and $p_i$ are the dimensionless area and perimeter of each cell, $\kappa_A$ is the rescaled area elasticity, and $p_0$ is the {\it preferred cell shape index} illustrating the cells homeostatic shape(see Methods)}. To probe tissue response, we applied quasi-static simple-shear deformation using Lees-Edwards boundary conditions, incrementally increasing strain with the FIRE algorithm to minimize energy (see Methods).

\begin{figure*}[htbp]
\includegraphics[trim=0 0 0 8cm, width=0.8\textwidth]{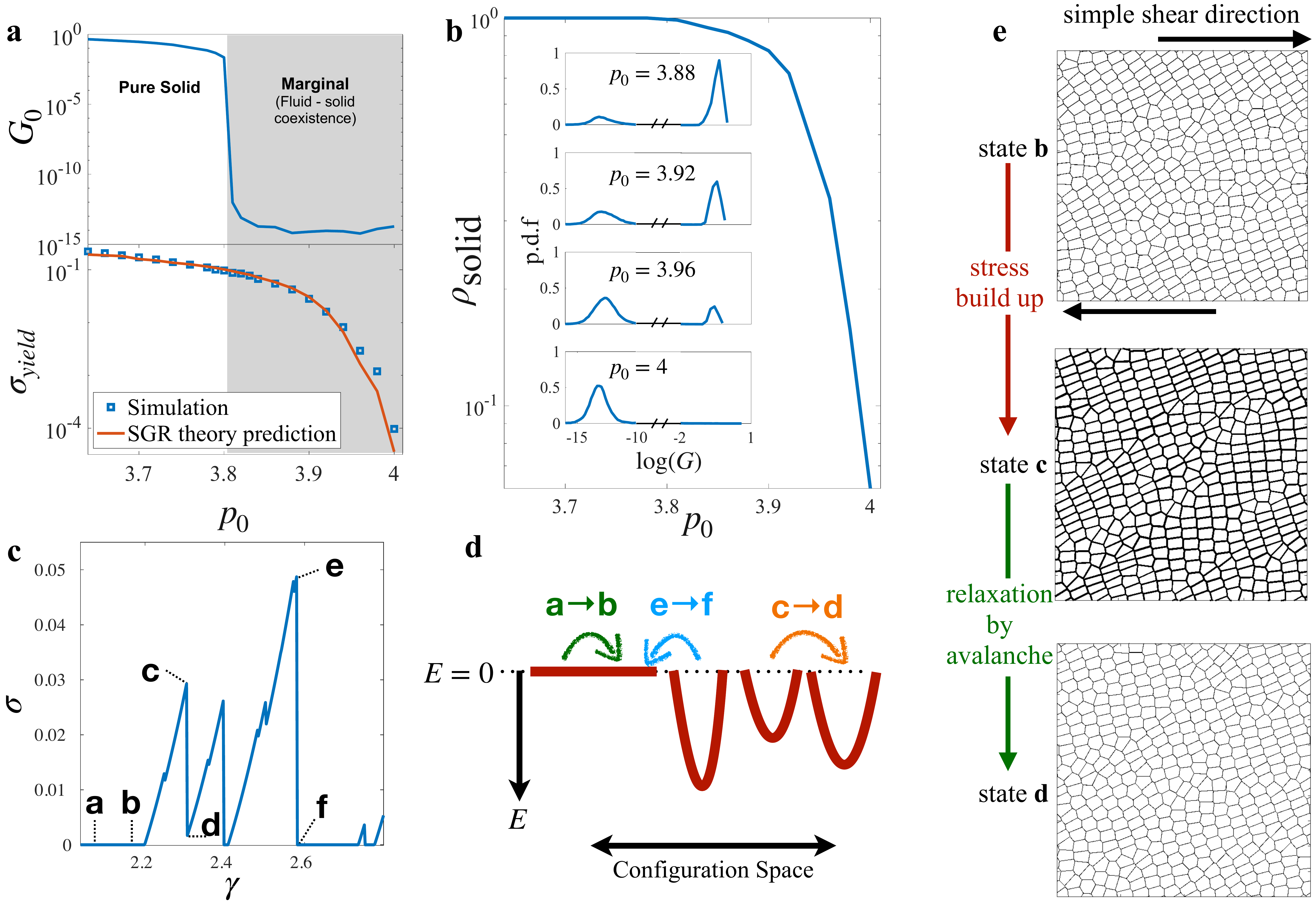}
\caption{
(\textbf{a}) Discrepancy between shear start-up and steady-shear regime. Top: The shear modulus of the un-sheared tissue ($\gamma=0$).  The shear modulus is obtained using linear-response calculation (see Methods). Bottom: The yield stress $\sigma_{yield}$ obtained  from the steady-state shear regime of quasi-static simulations is shown together with the yield stress obtained from the SGR model, where the only fitting parameter in the model, the elastic constant of an element was chosen to be $k=0.0386$.
(\textbf{b}) The probability of finding the system in solid state as a function of $p_0$. Inset: Distribution of tissue stress at different $p_0$. 
(\textbf{c}) Stress-strain curve example showing different yielding types: a fluid state yields to another fluid states \state{a} $\to$ \state{b}, a solid state yields to another solid one \state{c} $\to$ \state{d}, and a solid state yields to a fluid state \state{e} $\to$ \state{f}. 
(\textbf{d})  Schematic of the dynamics of elements in the SGR model: The energy landscape of the material consists of traps with different depth $E$ drawn from a distribution $\rho(E)$ that characterizes the structural disorder of the material. Yielding events are captured by transitions from one trap to another. The three types of transition illustrates the transitions observed in simulation.
{ (\textbf{e}) Simulation snapshots at different stages of a stress build-up and relaxation process. The edge thickness represents the edge tension, with thicker edges indicating higher tension. Here the states  correspond to the ones labeled in panel \textbf{e}.}}
\label{SGR}
\end{figure*}

In the absence of shear, it has been demonstrated that the preferred cell shape index $p_0$ drives a rigidity transition at $p_0 = p_0^* \approx 3.81$, where the linear response shear modulus vanishes  ~\cite{bi_nphys_2015}. Recent studies  ~\cite{huang2022shear,fielding_pre_2023} have shown that beyond this transition point, in the liquid phase ($p_0 > p_0^*$), the model can undergo strain-stiffening, indicating a rigidity gain upon strain application. In our quasi-static shearing protocol, we explore beyond the linear response and shear startup regimes into the large deformation {or steady-shear} limit . In this regime, the tissue exhibits plastic flow primarily through cell-cell rearrangements, or T1 transitions. {Here, the tissue’s yield stress is given by} 
\begin{equation}
\sigma_{yield} = \lim_{\dot{\gamma} \to 0} \langle \sigma(\dot \gamma) \rangle
\end{equation}
{where $\sigma$ is the $xy$-component of the stress tensor as we apply simple shear (see Methods for definition). The average is taken over all strain values in the steady shear regime.}
As illustrated in Fig.\ref{SGR}(a), while the startup shear modulus {$G_0 \equiv \lim_{\gamma \to 0} \partial \sigma / \partial \gamma$} computed using Eqn.\eqref{shear modulus} vanishes  ~\cite{huang2022shear} at the rigidity transition at $p_0 \approx 3.81$, signaling a solid-to-fluid transition, the yield stress $\sigma_{yield}$ does not disappear. Instead, it vanishes at a higher cell shape index, $p_0 \approx 4$. This  underscores a drastic difference in tissue responses between the transient shear startup and steady-shear regimes.

Under steady  shear and at shape indices higher than the rigidity transition associated with shear startup (i.e., at $p_0>p_0^*$), initially fluid-like systems can intermittently exhibit solid-like behavior before reverting to fluid-like states after yielding (Fig.\ref{SGR}(c)). {Here the solid-like states are chracterized by a non-zero shear modulus,(e.g., states \state{c, d, e} in Fig.\ref{SGR}c), while states that do not resist shear deformation, indicated by having zero shear modulus, are fluid-like (e.g., states \state{a, b, f} } in Fig.\ref{SGR}(c)). 
{To transition from state \state{b} to state \state{c}, the system gains rigidity under shear, moving from a fluid-like to a solid-like state. Stress then builds up, as indicated by edges experiencing high tension (Fig. \ref{SGR}e). Once sufficient stress is stored, the system yields, relaxing this stress via collective rearrangements, commonly referred to as avalanches, and transitions from state \state{c} to state \state{d}.}

{In the steady-shear regime, we can compute the instantaneous shear modulus (see Methods) to quantify solid vs. fluid-like states.}
The solid-fluid coexistence shows up as a bimodal distribution of the shear modulus, $p.d.f(G)$ shown in Fig. \ref{SGR}B, where the fluid phase is associated with a peak near the numerical noise floor of shear modulus ($\sim 10^{-12}$) while the solid phase corresponds to a finite shear modulus. The shifting behavior in the distributions can be quantified by the fraction of solid states $\rho_{solid}$ shown in Fig.\ref{SGR}(b). 
States below the rigidity transition $p_0 = p_0^* \approx 3.81$ therefore are always in the solid phase, which we term a {\bf pure solid}. In the range of   $p_0 \in [3.81,4]$,  $\rho_{solid}$ drops below $1$ indicating a solid-fluid coexistence, which we will refer to as {\bf marginal}. For $p_0 > 4$ the tissue remains always in the fluid phase as it cannot build up stresses in respond to shear strain. This is also consistent with the yield stress vanishing at $p_0 \simeq 4.0$.  
The fact that the material response depends on the application of shear is reminiscent of shear jamming in granular materials, where an state below the isotropic (un-sheared) jamming threshold can be jammed with the application of shear  ~\cite{bi2011jamming,behringer_bulbul2018physics,babu2022criticality,babu2023discontinuous}. The coexistence of solid and fluid phases also has analogs in  dense suspensions near shear jamming  ~\cite{shah2022coexistence} and discontinuous shear thickening  ~\cite{Morris_coexistence_review,hertaeg2024discontinuous}.

 \subsection{Predicting the tissue yield stress using a refined Soft Glassy Rheology model}
Given the continuous behavior of yield stress across the pure solid - marginal state transition, we aimed  to develop a unified model to deepen our qualitative understanding of the steady-shear regime properties using the Soft Glassy Rheology (SGR) framework   ~\cite{Sollich_PRL_1997,SGR_Sollich_1998}.
In the SGR model,  mesoscopic elements,  characterized by  elastic constant $k$ and local strain $l$, are confined within energy traps $E$, where they accumulate elastic energy as macroscopic strain increases, approaching a yield point either directly or through an activated "hop" driven by mechanical fluctuations from neighboring elements yielding. 
The material's dynamics under shear are governed by the  probability $P(E,l,t)$ {evolving in time $t$}, which follows the  Fokker-Planck equation  ~\cite{Sollich_PRL_1997,SGR_Sollich_1998}:
\begin{equation}
    \frac{\partial}{\partial{t}} P(E,l,t)=-\dot{\gamma}\frac{\partial P}{\partial l} - \Gamma_0 e^{[E-kl^2/2]/x}P+ \Gamma(t)\rho(E) \delta (l).
    \label{Sto_DE}
\end{equation}
The first term in Eq. \ref{Sto_DE} represents the motion of the elements driven by the applied shear rate, $\dot{\gamma}$. The second term describes thermally activated hopping from a trap with an effective depth of $E - k l^2/2$, which corresponds to the remaining distance to yielding. $x$ and {$\Gamma_0$} represents the mechanical noise in the system akin to temperature and {the hopping rate, respectively.
The final term represents the transition to a new trap with energy $E$ drawn from a quenched random distribution $\rho(E)$. The Dirac-delta function $\delta(l)$ reflects the assumption that the local strain $l$ is reset to zero after yielding. The total yielding rate at $t$,  $\Gamma(t)$, is explicitly defined in the SI text. }

In the SGR model, the choice of $\rho(E)$'s functional form critically influences material behavior  ~\cite{SGR_Sollich_1998}. Direct measurement of energy barrier distributions is challenging, leading prior studies to adopt generic or ad hoc assumptions for $\rho(E)$, such as exponential distributions  ~\cite{Yin_2008, SGR_Laponite_Bonn_2002, SGR_cytoskeleton_Mandapapu,bi2009rheology}. In this work, we introduce a novel approach based on distinct mesoscopic tissue phases observed: (1) fluid elements with zero yielding energy ($E = 0$) and (2) solid elements with finite yielding energy ($E>0$). Consequently, we propose a refined $\rho(E)$:
\begin{equation}
    \rho(E) = f_0\delta(E) + (1-f_0)\frac{E^{\kappa-1}e^{-E/x_0}}{\Gamma(\kappa) x_0^\kappa}.
    \label{rho_E}
\end{equation}
Here, $f_0$ denotes the probability of an element transitioning to a state with $E=0$, and $1-f_0$ corresponds to transitions into states with energy sampled from a k-gamma distribution, parameterized by mean $x_0$ and shape factor $\kappa$.
This is based on the previous observation that  the energy barriers to the T1 transition follow a k-gamma distribution  ~\cite{Bi_2014,bi_nphys_2015} with $\kappa \approx 2$. Together, Eqs. \ref{Sto_DE} and  \ref{rho_E}  describe three potential transitions in the energy landscape, depicted in Fig. \ref{SGR}: (1) fluid to fluid \state{a} $\to$ \state{b}, (2) solid to solid \state{c}$\to$ \state{d}, and (3) solid to fluid \state{e} $\to$ \state{f}.

We next examine the steady state behavior of Eq. \ref{Sto_DE} in the quasi-static limit ($\dot{\gamma} \to 0$), with details shown in the SI text.   The behavior is  governed by three parameters: the dimensionless ratio of mechanical noise to mean yielding energy $\chi = x/x_0$, the probability $f_0$ of transitioning to a fluid state, and the elastic constant $k$ of solid elements.

An important aspect of the SGR model is that the fluctuations that drive element rearrangements come from the mechanical noise due to other surrounding  rearrangement events in the system. 
These fluctuations are analogous to the energies released during yielding events observed in our simulations. To correlate this mechanical noise with our empirical data, we introduce the following relationship:
\begin{equation}
\chi = \frac{x}{x_0} \propto \frac{\langle \Delta E \rangle}{\langle E \rangle}.
\label{chi}
\end{equation} 

Here, $\langle \Delta E \rangle$ represents the average energy dissipated during yielding events, while $\langle E \rangle$ denotes the average energy of cells in their solid state.
Next, by analyzing the steady-state solution of Eq. \ref{Sto_DE}, we determine the probability that an element is in the solid phase as a function of $f_0$ (details in SI, Eq. \ref{fraction_and_f0}). This corresponds precisely to $\rho_{solid}$ in our simulations (Fig. \ref{SGR}B). Finally, we treat the elastic constant $k$ of the elements as a constant, independent of the shape index $p_0$. Given that both $\big<\Delta E\big>/\big<E\big>$ and $\rho_{solid}$ depend on $p_0$, the yield stress predicted by the SGR model (detailed in the SI) effectively varies only with $p_0$. This approach contrasts with previous studies that employed the SGR model  ~\cite{Yin_2008,SGR_Laponite_Bonn_2002,SGR_cytoskeleton_Mandapapu,samaniuk_emulsion_sgr}, where $\chi = x/x_0$ was often used as a fitting parameter. In our research, we derive $\chi$ directly from simulation data, enhancing the predictive accuracy of our theoretical results and distinguishing our use of the SGR model as predictive rather than merely descriptive. In Fig. \ref{SGR}(a), we plot the SGR-predicted yield stress as a function of $p_0$. This demonstrates that the SGR model accurately predicts the yield stress vanishing point and its dependence on the cell shape index $\sigma_{yield} (p_0)$.

The dual-state SGR model identifies two primary mechanisms responsible for the yield stress transition:
{\bf (1)} As $p_0$ increases, states with zero yielding energy barriers become more prevalent, leading to frequent yielding under deformation. This behavior is depicted by transitions such as \state{a} $\to$ \state{b} and \state{e} $\to$ \state{f} in Fig. \ref{SGR}d; 
{\bf (2)} Concurrently, mechanical noise from stress redistributions approaches the scale of the yielding energy barriers, enhancing the likelihood of solid-solid transitions through activated processes induced by neighboring rearrangements, as illustrated by \state{c} $\to$ \state{d} and \state{e} $\to$ \state{f}.

 \subsection{{A novel method to infer tissue stresses based on instantaneous snapshots.}}

{ The coexistence of phases observed in the Vertex-based model reinforces the idea that tissues function as yield stress materials while also displaying fluid-like behavior. When combined with Soft Glassy Rheology (SGR) theory, this coexistence provides predictions for tissue yield stress based on the homeostatic cell shape index $p_0$. Given $p_0$, the Vertex based model could also provide estimation of the instantaneous tissue stress using Eqn. \ref{def:stress}. However, determining the homeostatic-target shape index $p_0$
remains an experimental challenge. In contrast, segmented cell configurations in tissues are experimentally accessible and have been utilized in several non-invasive stress inference methods such as Bayesian Inference method  ~\cite{Chiou_2012,Ishihara_2012} and the image-based Variational method  ~\cite{Noll_Shraiman2020}. Although the Bayesian Stress Inference method proposed by Ishihara and Sugimura  ~\cite{Ishihara_2012} has been applied to various systems, including Drosophila notum, retinal ommatidia, germband, and the quail early embryo  ~\cite{ishihara2013comparative,kong2019experimental}, its results strongly depend on a prior distribution of edge tension and cell pressure, which is not necessarily normally distributed, as originally proposed. Additionally, the method faces the challenge of having more unknowns than constraints. In contrast, the Variational Method proposed by Noll et al.  ~\cite{Noll_Shraiman2020} addresses the issue of underconstrained variables but relies on a computationally expensive fitting approach. Here, we propose a fast, non-invasive, image-based tissue stress inference method that is both convenient and accurate. It has been shown that edge length distribution is closely related to the distance to the yield stress in monolayers  ~\cite{Popovic_2021}. However, the connection between edge length distribution and tissue stress has not been quantified. In this work, we demonstrate that the cumulative distribution of edge length elegantly serves as a robust estimator of tissue stress.}

\begin{figure*}[htbp]
\includegraphics[width=0.8\textwidth]{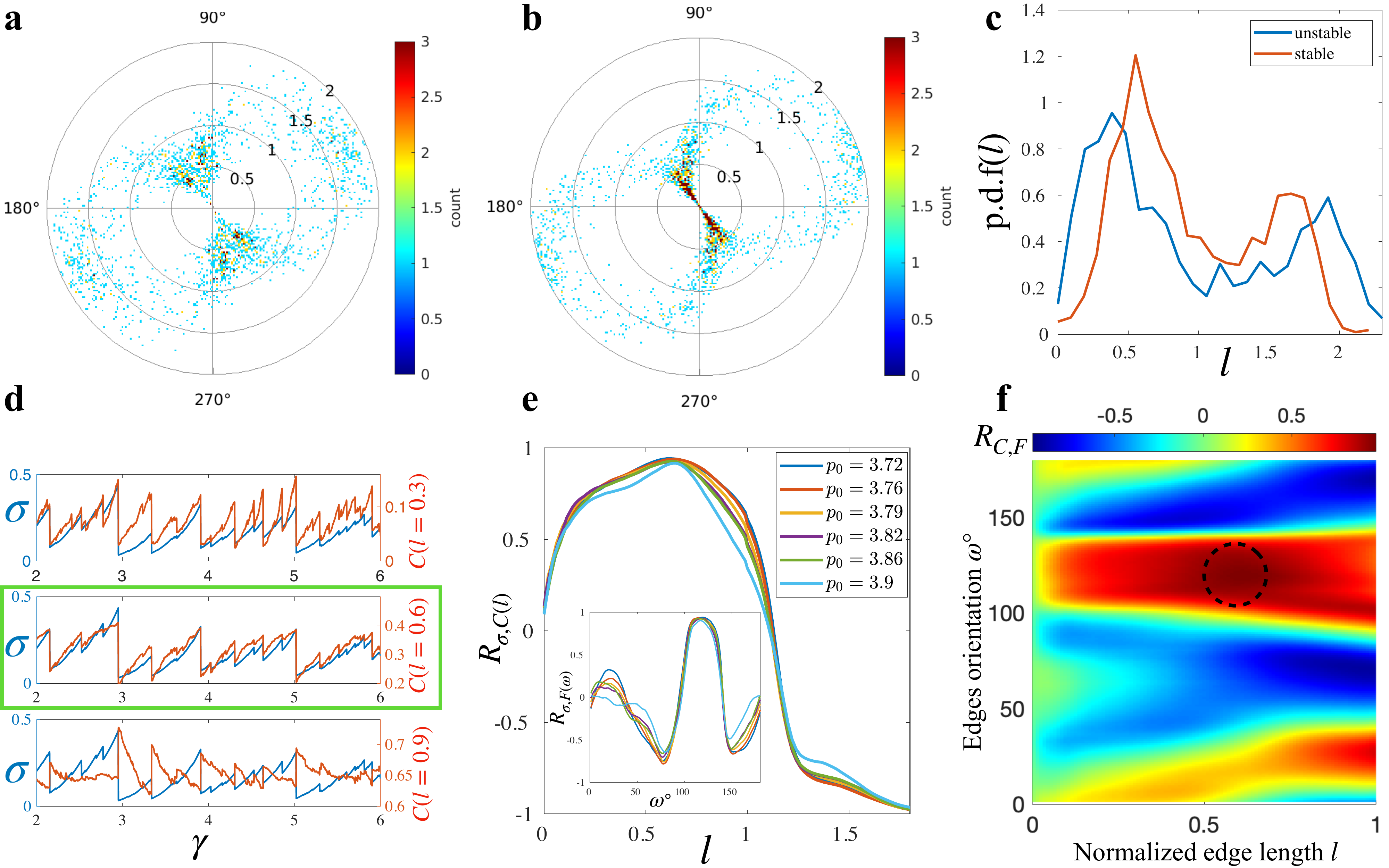}
\caption{{A new approach to infer tissue stress: (\textbf{a}) Polar distribution of cell edge vectors in a stable state. (\textbf{b}) Polar distribution of cell edge vectors in an unstable state. (\textbf{c}) Edge length density distribution at different instability. Panels (a-c) were created using $p_0=3.72$. (\textbf{d}) The stress-strain curve is overlaid with $C(l)$-strain curve at different $l$. The green box denotes the trace with high correlation between $\sigma$ and $C(l)$. (\textbf{e}) The correlation between tissue stress and the cumulative edge length distribution $R_{\sigma, C(l)}$ suggests a robust critical normalized edge length $l^*$. Inset: The correlation between tissue stress and the portion of edges at different orientation $R_{\sigma,F(\omega)}$. (\textbf{f}) Color-plot of the correlation $R_{C,F}$ between edges length distribution and edges orientation distribution. The black-dashed circle indicates the region in where the correlation is maximized, from which $l^*$ could be determined.}}
\label{stress inference}
\end{figure*}

{In systems under deformation, the configuration exhibits anisotropy, with preferred orientations dictated by the deformation direction. To analyze this anisotropy and its strain evolution, we examined the distribution of edge vectors in the system, represented as polar distributions. In these distributions, color encodes the frequency of edges with lengths and orientations defined by radial distance and azimuthal angle, respectively. The polar distribution for an isotropic system therefore looks like a circle with dots scattered randomly inside. This is not what we observed in our system due to the anisotropy. To systematically study across different systems and time frames, we used the normalized edge length $l=L/\bar{L}$, obtained by normalizing the edge lengths by the average edge length in the current snapshot. Fig.\ref{stress inference}a displays the polar distribution of the normalized edge vectors $\overrightarrow{l}$ in a low-stress, stable system far from yielding. Conversely, Fig.\ref{stress inference}b depicts the edge vector distribution in a high-stress, unstable system close to yielding. In both cases, anisotropy emerges, with longer edges aligning with the shear direction (about 45 degrees with respect to horizontal) and shorter edges perpendicular to it. However, the trend is more pronounced in unstable systems, illustrated by the red color band representing high population of short edges with orientation about 135 degrees with respect to horizontal (Fig.\ref{stress inference}b), implying a connection between instability and edge vector distribution. Furthermore, these polar plots reveal a strong dependence between edge length and orientation in deformed configurations.}

{The correlation between edge length and orientation suggests that the edge length distribution can effectively represent the overall edge vector distribution. To investigate how edge configuration influences instability, we compared the edge length distributions of stable and unstable states. As illustrated in Fig.\ref{stress inference}c, both distributions exhibit a bimodal shape due to anisotropy. However, in the distribution of unstable system, the separation between the two peaks is larger, reflecting stronger anisotropic effects and an abundance of short edges, which are more prevalent in the unstable configuration. This motivated us to study the evolution of the edge length distribution as strain increase.}

{Investigating the evolution of the edge length cumulative distribution $C(l)$ in our quasi-static simple shear simulations, we observed a correlation between $C(l)$ and $\sigma$, with the correlation level depending on $l$. As shown in Fig.\ref{stress inference}d, while the correlation $R_{\sigma,C}$ between tissue stress $\sigma$ and $C(l)$ varies with $l$, there exists a range of $l$ where $C(l)$ is highly correlated with $\sigma$. We denote the $l$ value that maximizes the correlation $R_{\sigma, C}$ as $l^*$, and the corresponding cumulative distribution $C(l^*)$ as $C^*$. Interestingly, both $l^*$ and $R_{\sigma, C}$ remain robust with changes in $p_0$, with the critical correlation $R_{\sigma, C^*}$ exceeding 0.9, as shown in Fig.\ref{stress inference}e. Given the strong correlation between $C^*$ and $\sigma$, as well as the fact that cell edge lengths can be directly extracted from imaging, $C^*$ could serve as a non-invasive metric for inferring tissue-level stress.}

{While $l^* \approx 0.61$ is consistent across various $p_0$ values in our quasi-static simulations, its practical value may depend on system properties and the shearing direction. Thus, developing a metric to experimentally determine $l^*$ is crucial for the stress inference method. Recognizing the strong relationship between edge length and orientation in anisotropic systems, we investigated tissue stress and edge orientation correlations. To quantify the edge orientation distribution, we used the proportion of edges a given orientation $\omega$:
\begin{equation*}
    F(\omega) = \frac{N(\omega - \Delta \omega,\omega+\Delta \omega)}{N_{\text{edges}}}
\end{equation*}
Here, $N(\omega - \Delta \omega,\omega+\Delta \omega)$ represent the number of edges in a cone with the open angle of $2\Delta \omega$ at the $\omega$ direction and $N_{\text{edges}}$ is the total number of edges in the system. In this section, all angular values are in degrees. Tracking the evolution of $F(\omega)$ at different $\omega$ as strain increased, we observed a critical orientation $\omega^* \approx 120^\circ$, where $F\omega$ was highly correlated with tissue stress, with a correlation coefficient $R_{\sigma,F(\omega^*)}>0.9$ (Fig.\ref{stress inference}e inset). Although $R_{\sigma,F}$ depends on the cone wideness $\Delta \omega$, which we chose as $\Delta \omega =18^\circ=\pi/10$ to optimize $R_{\sigma,F}$, the critical orientation $\omega^*$ is independent of $\Delta \omega$. Additionally, $\omega^*$ was robust across variations in $p_0$ (Fig.\ref{stress inference}e inset). Given the high correlations $R_{\sigma,C}$ and $R_{\sigma,F}$, the critical edge length $l^*$ could potentially be determined from the correlation between edge length and edge orientation, $R_{C,F}$. As suggested by Fig.\ref{stress inference}e, we only focus on the range $l<1$ in which $C(l)$ is known to positively correlates with tissue stress $\sigma$. By computing the correlation $R_{C(l),F(\omega)}$ at different $(l,\omega)$, we identified a region where the correlation is exceptionally high (greater than 0.93), as indicated by the dashed circle in Fig.\ref{stress inference}f. The $l$ values in this region align excellently with the critical edge length $l^*$, suggesting that $l^*$ can be estimated as the normalized edge length value $l$ maximizing $R_{C(l),F(\omega)}$. With $l^*$ in hand, ones could use $C^*=C(l^*)$ to extract an estimation of the tissue stress as it evolves during the deformation process. Our stress inference method could be conveniently validated by experiments such as the monolayer stretching protocol  ~\cite{monolayer_stretching}. Although $C^*$ could be use directly as the inference for tissue stress, more complicated functional models are available and discussed in more detail in the SI. }

 \subsection{Dynamics of tissue plasticity}

\begin{figure*}[htbp]
\includegraphics[width=0.8\textwidth]{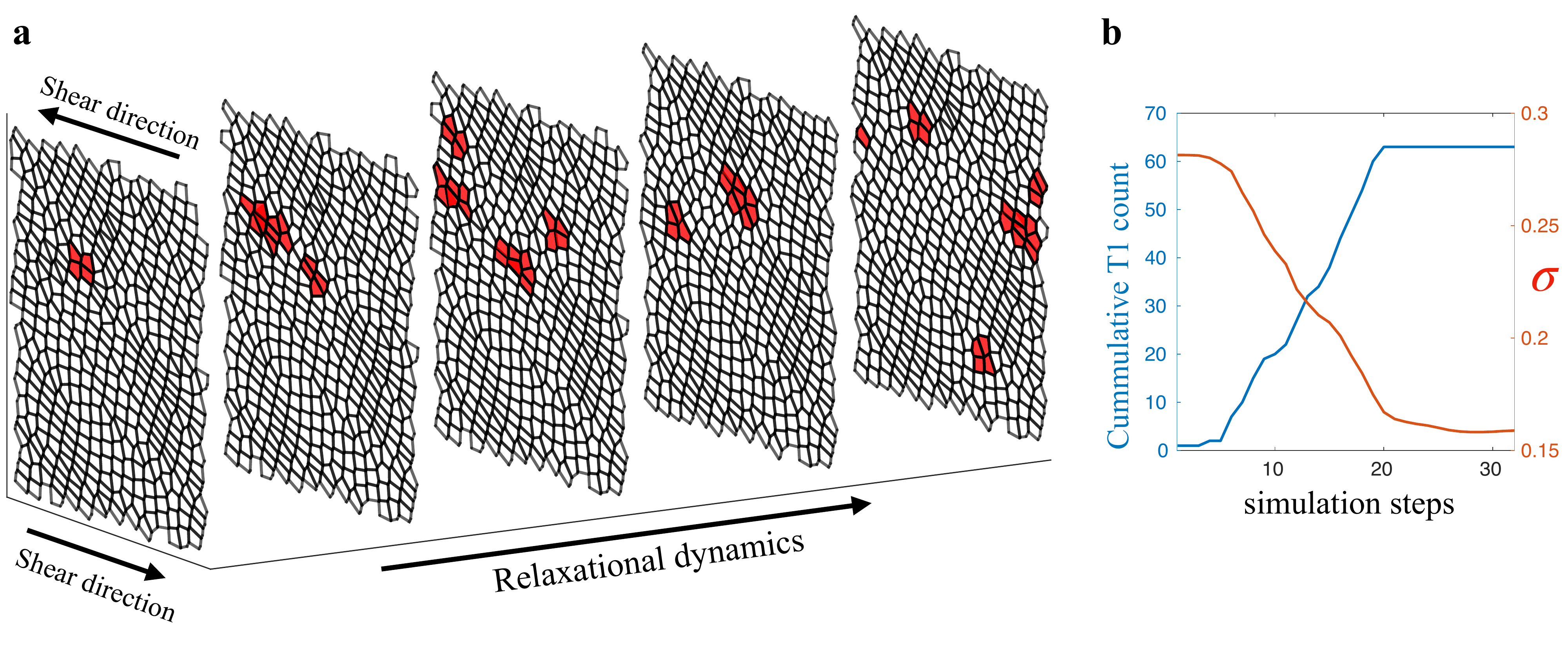}
\caption{Spatiotemporal map of plastic events in tissue. (\textbf{a}) Space-time map of plastics rearrangement in the form of T1 transition during an avalanche. Cells that participated in T1 transitions are labeled in red. The example avalanche was selected from a system at $p_0=3.72$. (\textbf{b}) Number of accumulative plastic rearrangement and tissue shear stress as the avalanche progresses.}
\label{per homo}
\end{figure*}

So far, we know that for a system in the coexistence phase to transitions from solid to fluid, a collective rearrangement event is required to significantly remodel the configuration. However, the mechanism that govern the occurrences of these events and the evolution of the system during the events is still not fully understood.
To explore the yielding behavior of biological tissues, it is essential to describe the dynamics of plastic events during avalanches. In this study, we examine the plasticity dynamics by investigating the spatiotemporal evolution of the plastic rearrangements created by other rearrangements during an avalanche ~\cite{Stanifer_LisaManning2022,richard2023mechanical}. Fig.\ref{per homo}a displays a space-time plot of the occurrences of T1 transitions during an avalanche, with the cells labeled in red indicating participation in the T1 transitions.

As depicted in Fig.\ref{per homo}a, an avalanche involving numerous plastic rearrangements could originate from a single T1 transition, which we refer to as the initial trigger. Starting from the initial trigger, the stress redistribution from each event can also stimulate surrounding cells to become unstable and undergo T1 transitions. {These {soft} cells, whose mechanical characteristics are distinct from the surrounding cells, experience larger deformation than the neighboring cells and have been observed in multicellular tumor spheroids using 3D light microscopy ~\cite{jaiswal2019elastography}. From amorphous solid point of view, these cells are referred to as soft spots  ~\cite{falk2011deformation,Manning_Liu_2011,Patinet_Falk_2016,yang2024multicell} or Shear Transformation Zones  ~\cite{Falk_1998,Manning_2007}.} This cascade of cellular rearrangements can therefore lead to an avalanche, which continues until the population of soft spots are sufficiently depleted. In Fig.\ref{per homo}b, we show the number of accumulated T1 transitions and the tissue shear stress during a typical avalanche. The stress relaxation due to an avalanche therefore is the origin of the discontinuous yielding of stress during quasistatic shear.  
\begin{figure*}[htbp]
\includegraphics[width=0.8\textwidth]{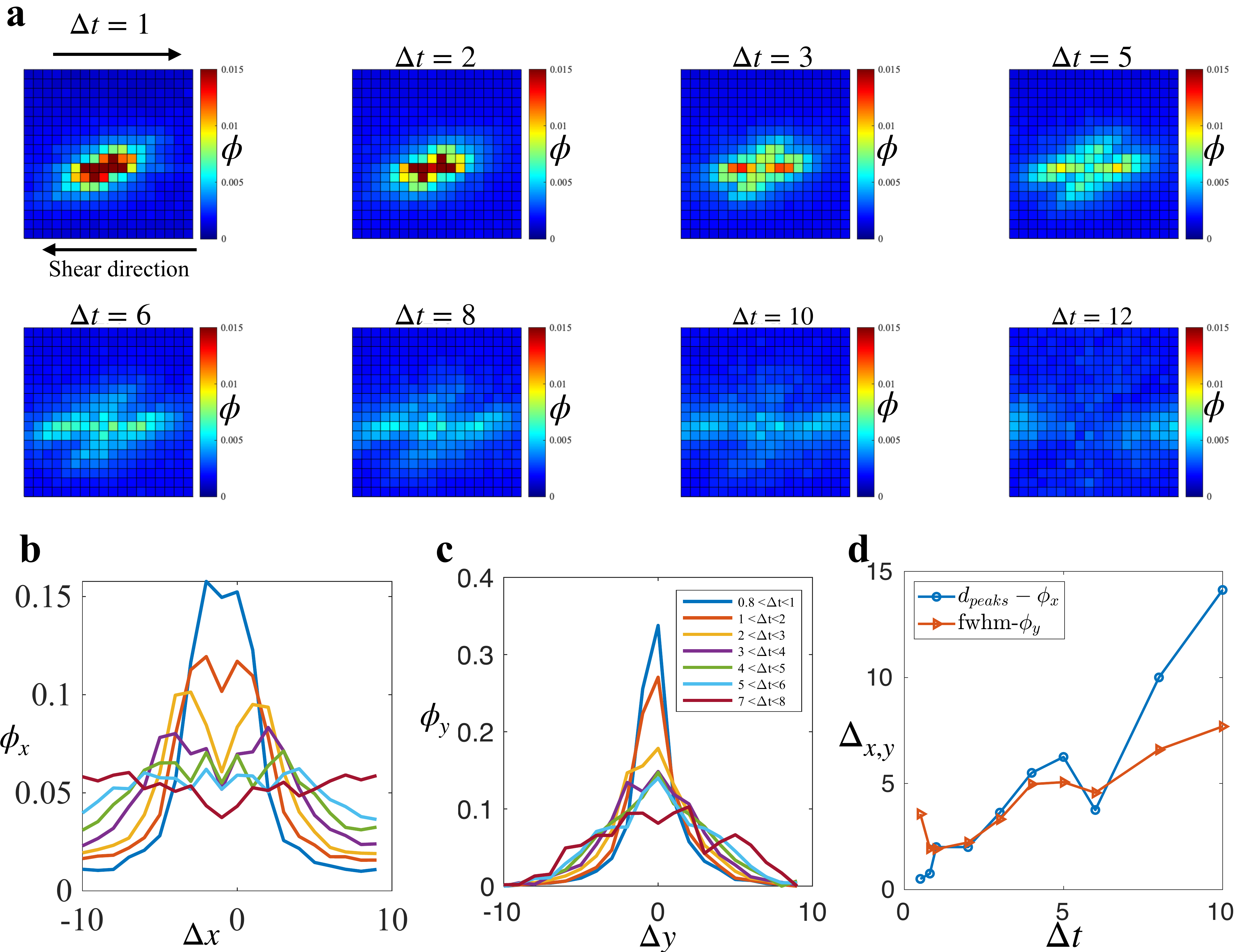}
\caption{Propagation of plastic events.
(\textbf{a}) Evolution of the probability field $\phi$ at $p_0=3.72$. Each image corresponds to the probability field at a particular time lag $\Delta t$. Bright regions indicate a high probability of finding another rearrangement in the region relative to the causal rearrangement.
(\textbf{b}) Spatial distribution of the correlation field $\phi_x$ as a function of the relative horizontal position $\Delta x$. The distribution has a diffusing bimodal shape, indicating a convection process alongside diffusion.
(\textbf{c}) Spatial distribution of the correlation field $\phi_y$ as a function of the relative vertical position $\Delta y$. The distribution is bell-shaped, with the width of the bell increasing as time progresses, indicating a pure diffusion process.
(\textbf{d}) The separation of the peaks in $\phi_x$ and the Full Width at Half Maximum (FWHM) of $\phi_y$ as functions of the time lag $\Delta t$.}
\label{stress propagation}
\end{figure*}
In Fig.\ref{per homo}a, the location of rearrangments over time suggests that there is a preferred direction for the avalanche to propagate. In order to quantify this and to establish a causal relationship in time, we define a two-point, two-time correlation function:
\be
\phi( \mathbf{r},\Delta t)=\langle P(\mathbf{r}_0,t_0)P(\mathbf{r}_0+\mathbf{r},t_0+\Delta t{)}\rangle,
\ee
where $P(\mathbf{r},t)$ is a binary field, representing the occurrence of a T1 transition (1 if  a T1 transition at occurs at $\mathbf{r}$ and time $t$ and 0 otherwise).  $\langle ...\rangle$ represents spatial and ensemble averaging.  
With this definition, $\phi$ is therefore the conditional probability of observing a T1 transition at location $\mathbf{r}_0+\mathbf{r}$ and time $t_0 + \Delta t$, given that a transition has already occurred at ($\mathbf{r}_0,t_0$).

In Fig.\ref{stress propagation}a, we illustrate the evolution of the field $\phi$ as $\Delta t$ increases. The field $\phi$ behaves like a wave that propagates away from the causal rearrangement and prefers to propagate in the x-direction, coinciding with the direction of the external shear force. The evolution of the field $\phi$ reflects the interplay between the stress redistribution from a plastic rearrangement and the population of soft spots which could rearrange under the effect of the strain field.

We first focus on the angular dependence of $\phi$, which  shows an anisotropic four-fold  pattern. This anisotropy is consistent with the stress redistribution field due to a rearrangement in an elastic medium as predicted by  elastoplastic models  ~\cite{picard2004elastoplastic, Popovic_2021}. However, it differs from the isotropic probability field observed in ductile, soft disk systems  ~\cite{zhang2021interplay}. This contradiction likely arises from the difference in shape anisotropy between soft disks and cells in our system. While soft disk systems exhibit minimal particle shape anisotropy, cells in our system can sustain large deformations and have highly anisotropic shapes. Consequently, deviatoric strain triggers rearrangements in soft disk systems, whereas simple shear strain is responsible for triggering rearrangements in our system.

To better understand how the rearrangement probability field propagates, we looked at $\phi_x = \frac{\phi(x,y=0)}{\sum_x \phi(x,y=0)}$ and $\phi_y=\frac{\phi(x=0,y)}{\sum_y \phi(x=0,y)}$ separately. $\phi_x$, as observed in Fig.\ref{stress propagation}b, is a bimodal distribution that evolves in time such that the distance between the two peaks $d_{peaks}$ increases as time goes. The diffusing bimodal distribution suggests that the propagation is a combination of convection and diffusion, and the drift velocity could be captured by the rate at which the peaks' separation increases. Conversely, $\phi_y$ is a bell-like shape distribution that gets broader as time progresses (Fig.\ref{stress propagation}c), indicating that the propagation in the y-direction is similar to a purely diffusion process with the diffusivity can be captured by the evolution of the FWHM. Fig.\ref{stress propagation}d shows that $d_{peaks}$ increases faster than the FWHM-$\phi_y$. Since the rearrangement probability field is the result of the shear stress redistribution and the population of soft spots, the propagating mechanism of the field also should agree with the propagation of the stress redistribution.

The phenomenon that the shear stress redistribution tends to propagate in the direction of shear has also been observed in particle-based systems governed by inverse-power-law pairwise potentials  ~\cite{richard2023mechanical}. Remarkably, this behavior bears a striking resemblance to the propagation of elastic waves. In the theory of elasticity, longitudinal waves, characterized by displacement in the direction of propagation, outpace transverse waves  ~\cite{landau2012_elastic_theory}. Moreover, longitudinal elastic waves involve changes in local density ~\cite{landau2012_elastic_theory}, akin to the x-propagating excitation wave's modulation of local density via T1 transitions. Conversely, transverse elastic waves do not induce density changes, resembling the infrequent involvement of T1 transitions in y-propagating excitation waves. 
Notably, the mechanism driving stress redistribution to preferentially propagate in the shear direction appears universal, independent of $p_0$. However, since $p_0$ governs the elastic
response in our system, with higher values corresponding to a less elastic state, there is a negative correlation between the stress redistribution wave's speed and $p_0$.

\begin{figure}[htpb]
\includegraphics[width=\columnwidth]{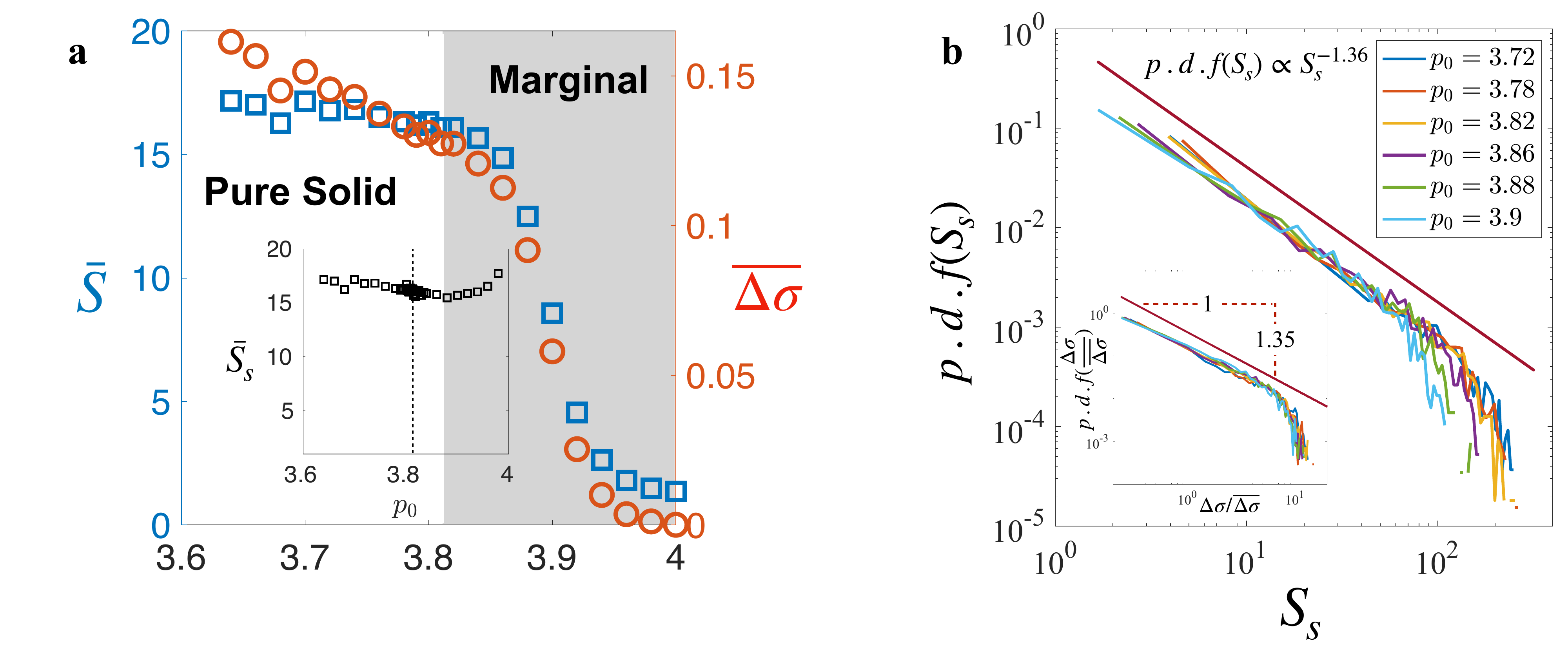}
\caption{Avalanches statistics. 
(\textbf{a}) Dependence of average yielding size $\bar{S}$ and average stress relaxed $\overline{\Delta \sigma}$ on $p_0$. Inset: Dependence of average avalanches size $\bar{S_s}$ on $p_0$.  
(\textbf{b}) Distribution of avalanche size follows a power-law distribution. Inset: Distribution of scaled average stress relaxed by avalanches also follows the same power law}
\label{avalanches size stats}
\end{figure}

 \subsection{Statistics of tissue avalanches} In addition to the universal propagation mechanism, we wondered if the statistics of tissue yielding events also exhibit universality. In Fig.\ref{avalanches size stats}a, both the average yielding size $\bar{S}$, denoting the total number of T1 transitions after a yielding event, and the average stress drop, representing the amount of stress relaxed by the event, exhibit the same dependence on $p_0$. In the solid regime, while the stress decreases with increasing $p_0$, the average yielding size shows minimal variation. This trend of $\bar{S}$ versus $p_0$ is akin to that observed in Fig.\ref{SGR}b for the proportion of the solid state.

However, in the marginal phase, there are different types of yielding events as discussed previously (Fig.\ref{SGR}{c}). In the yielding events that occur while the system is fluid-like, illustrated by the \state{a} $\to$ \state{b} transition in Fig.\ref{SGR}{c} (Type \cf{I}), the tissue lacks rigidity and therefore is unable to transmit stress to initiate a cascade of rearrangements. Conversely, yielding events following a solid state, illustrated by \state{c} $\to$ \state{d} and \state{e} $\to$ \state{f} transitions (Type \cf{II}), tend to be cascading as the rigid tissue is capable of propagating the stress redistribution. It is this latter type that we refer to as tissue avalanches from now on. Since the avalanches growing mechanism is universal, we expect their statistics to also be independent of $p_0$.

Excluding yielding events of type \cf{I} from the analysis and specifically analyzing only the avalanches, we indeed find that the average avalanche size $\overline{S_s}$ does not vary significantly with $p_0$ (Fig.\ref{avalanches size stats}a inset), suggesting  {universal avalanches size statistics} . To rigorously assess this universality, we examine the distribution of avalanche sizes across various $p_0$ values (see Fig.\ref{avalanches size stats}b). Strikingly, we observe a consistent power-law distribution, reminiscent of the Gutenberg–Richter law observed in earthquakes  ~\cite{Kawamura_2012,gutenberg1944frequency}, with an exponent $\tau = -1.36$, which agrees with the analogous exponent observed in overdamped elastoplastic models under shear  ~\cite{talamali2011avalanches,karimi2017amorphous_avalanche} and in vertex model on spherical surface ~\cite{amiri2023_avalanche_spherical_surface}. This shared characteristic suggests a parallel between biological and seismic avalanches and supports the argument that the vertex model and epithelial tissues belong to the universality class of plastic amorphous systems.

Furthermore, we find that the same power law applies to the distribution of average stress drops during avalanches when scaled by the average (Fig.\ref{avalanches size stats}b inset). This collapse after rescaling implies that the stress relaxation mechanism via avalanches is independent of the shape index $p_0$, and $p_0$ only affects the average stress relaxed by governing tissue overall stiffness. Moreover, the similarity in the stress drop distribution and avalanches size distribution indicates that each plastic rearrangement, on average, releases a similar amount of stress that depends only on $p_0$. The convergence of these distributions suggests that the growth of avalanches remains unaffected by changes in $p_0$, providing additional evidence for the universal propagation mechanism discussed earlier.

 \subsection{Predicting tissue avalanches based on static structural information} 
While the first cells to undergo a T1 transition triggers the avalanche, in order for the avalanche to grow, it is necessary to have soft spots in the system that are susceptible to undergo T1s. 
In the framework of the elastoplastic model, it has been established that the distance to yield $x$, which represents the additional stress required to trigger a yielding event, follows a power-law distribution, $p(x) \propto x^{\theta}$  ~\cite{Lin_Wyart_2016,Lin_2014,Lin_EPL}. The exponent $\theta$ has been suggested as a measure of the system's instability, with a higher value indicating a more stable state.

In the vertex-based model family, it has been proposed that the distance to yield $x$ exhibits a linear relationship with the length of cell edges $L$, and that the distribution of edge lengths should follow the same power-law behavior as $\rho(x)$  ~\cite{Popovic_2021}. While this argument establishes a connection between system configuration and instability, the efficacy of using the distribution of short edges to describe instability remains uncertain.

\begin{figure}[htpb]
\includegraphics[width=1\columnwidth]{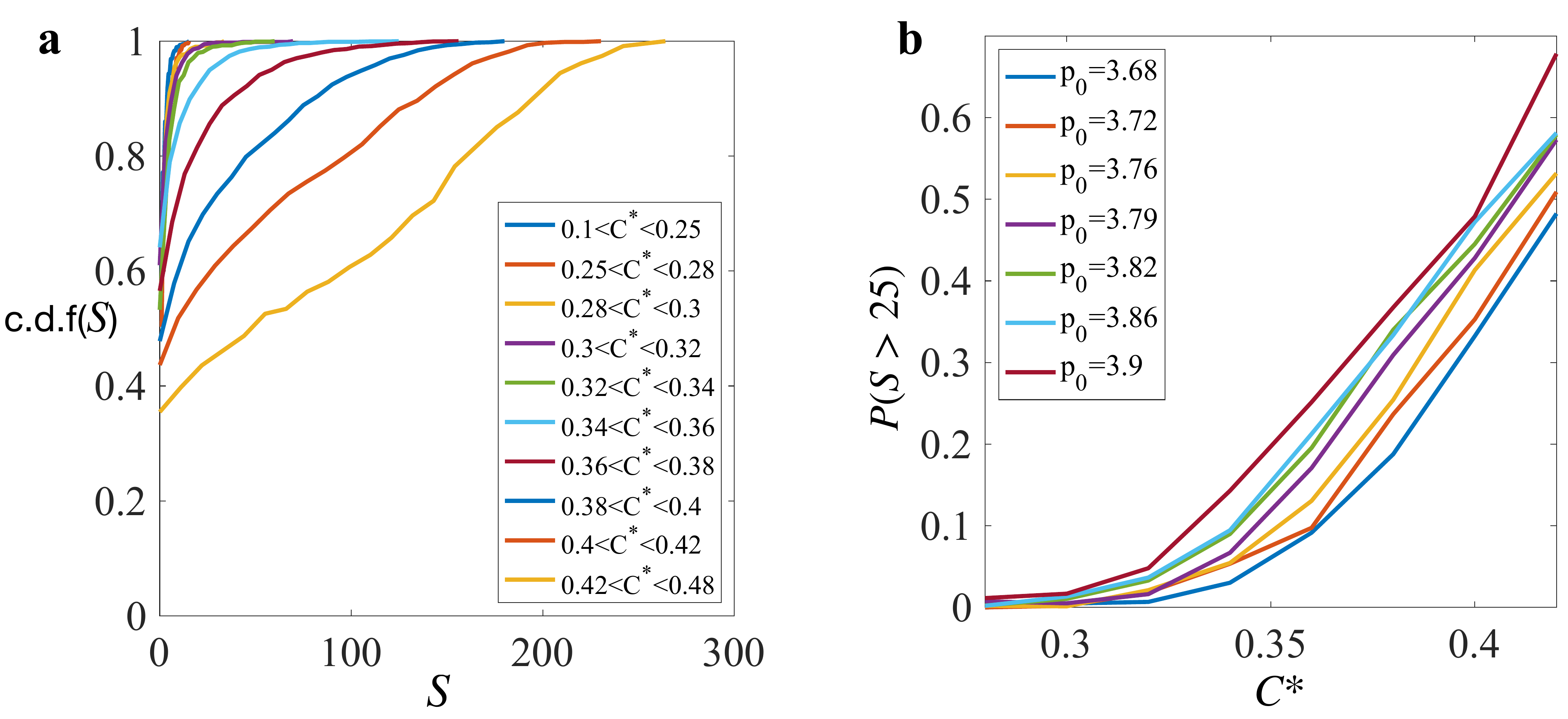}
\caption{Predicting avalanches. 
(\textbf{b})An example of the evolution of tissue shear stress $\sigma_{xy}$ and $C^*$ as strain increases.  
(\textbf{b}) Scatter plot of shear stress and $C^*$ at different $p_0$ in the solid regime. 
(\textbf{c}) c.d.f($S$) at different $C^*$ level at $p_0=3.79$. 
(\textbf{d}) Probability of having avalanches of size greater than 25 at different $C^*$}
\label{C as order param}
\end{figure}

To address this ambiguity, we investigate this concept within our Voronoi model and observed an intriguing correlation between the exponent $\theta$ and system instability (see S.I and Fig.\ref{theta exponent}). However, $\theta$ is not a reliable metric because it is derived from a power-law fit that heavily depends on the range of fitting  ~\cite{Popovic_2021}. In practice, the cumulative distribution function (c.d.f) of $L$ exhibits power-law behavior only within a specific range, which varies from sample to sample. Therefore, we propose using {$C^*$} as the parameter of instability to avoid the uncertainties and biases associated with fitting. 

Using $C^*$ as the parameter of instability, with higher $C^*$ corresponding to a more unstable state, we observed a relationship between instability and avalanche size. {To quantify the relation between instability and avalanche size, we focus on the failure states, i.e states that are about to yield, indicated by having a stress drop in the next strain step and a corresponding avalanche.} We computed $C^*$ for those states and then grouped the avalanche sizes based on $C^*$ (Fig.\ref{C as order param}a). At low $C^*$, if the tissue yields, the size of the yielding event is likely to be small, indicated by a rapid increase to 1 in the cumulative distribution function (c.d.f.) of $S$. As $C^*$ increases, the likelihood of larger yield events grows. The probability of observing an avalanche of size 25 or greater is summarized in Fig.\ref{C as order param}b.

While $C^*$ can provide predictions about avalanche size, it cannot determine when an avalanche will occur. Therefore, we require another tool to forecast yielding events. In amorphous solids, the locations of plastic rearrangements during an avalanche largely depend on the material's structural configuration, with areas more likely to experience plastic events called soft spots. Various frameworks have been proposed to link structure and plasticity, such as the Shear Transformation Zone theory  ~\cite{Falk_1998,Patinet_Falk_2016,Richard_2021,LANGER_2006,Manning_2007} and lattice-based models. The most promising theoretical approach for predicting the locations of soft spots involves identifying these areas based on soft vibrational modes  ~\cite{Manning_Liu_2011,rottler2014predicting,ding2014soft} . As a system approaches a plastic rearrangement, at least one normal mode  is supposed to approach zero frequency  ~\cite{Manning_Liu_2011} . However, the vibrational mode analysis is not applicable to the vertex-based model family due to the cuspiness of the energy landscape  ~\cite{Bi_2014,bi_nphys_2015} . In such systems, the energy cusp at a plastic event prevents the corresponding low-frequency mode from vanishing as it would in systems with smooth, analytic energy landscape. As shown in Fig.\ref{normal modes}, in our system, no low-frequency mode approaches zero frequency except at the onset of the plastic event. Hence, an alternative approach is necessary.

In our model, the deformation of edges immediately following shear strain is deterministic. Through simple geometry, we deduce the existence of a range of orientations wherein edges are prone to shortening upon shearing, rendering them more susceptible. In the vertex model, under the condition $\dot{\gamma} \ll 1$, the change in edge length $\delta L$ due to shear is approximated as $\delta L \approx \dot{\gamma} L \sin(2\phi)$, where $L$ and $\phi$ represent the edge length and orientation, respectively. Consequently, in the vertex model, the most {susceptible} orientation is $\frac{3\pi}{4}$. If an edge is sufficiently weak (or short) and happens to align with this susceptible orientation, it may yield under the influence of shearing, potentially triggering further rearrangements in its vicinity. We refer to these {susceptible} edges as triggers. Since triggers are local elements, their presence is not captured by the cumulative distribution function of edge lengths, c.d.f($L$), thereby explaining why $C^*$ alone cannot predict imminent system failure. In summary, the presence of a trigger is the necessary condition, while a high $C^*$ in the current state is the sufficient condition for a large avalanche in a tissue monolayer.

\section*{Discussion}

We studied the response of tissue monolayers to external quasi-static shear stress in the long-term steady-shear regime. For a tissue monolayer initially in a fluid state, it behaves like a yield-stress material in the shear buildup regime but eventually enters a marginalphase in the long-term steady shear regime. Incorporating the coexistence phase into to the SGR model, we elucidated the discrepancy between the rigidity transition and the yield stress transition observed in the Vertex model under simple shear. 

Besides the yield stress, tissue plasticity is also reflected in avalanches of plastic rearrangements. {Understanding those local rearrangements would help reveal the mechanism of mechanical homeostasis and reparation in biological tissue}. By studying the dynamics of tissue avalanches, we observed a universal propagation mechanism of plastic events that is independent of the shape index $p_0$ and has two preferred directions, with the direction of the external shear being the one with faster propagation. The rearrangement probability $\phi$ studied here is closely related to, but not identical with, the stress redistribution predicted by elastoplastic models  ~\cite{picard2004elastoplastic,Popovic_2021} or the strain field due to rearrangement  ~\cite{zhang2021interplay}. Instead, $\phi$ captured the interplay between stress redistribution and spatial distribution of weak spots. Since most edges that participated in T1 transition in our system orient at $-\pi/4$ with respect to the horizontal (Fig.\ref{T1 orient}), the positive shear stress is expected to symmetrically redistribute along the horizontal and vertical directions  ~\cite{Popovic_2021}. Although $\phi$ does propagate mainly in the horizontal and vertical directions, it does not possess the vertical-horizontal symmetry seen in stress redistribution or in the strain field. This difference arises from highly heterogeneity and anisotropic in the spatial distribution of weak spots. {The rearrangement dynamics studied in this work could help identify regions where rearrangements are anticipated in the near future. Predictions based on our analysis are most reliable in systems where external forces dominate cellular activity or under quasi-static conditions. In these scenarios, rearrangements occur systematically, driven solely by stress redistribution from regions of high stress to relax global stress. Conversely, in systems with high cellular activity, rearrangements happen randomly due to the stochastic motion of cells, blurring the impact of stress redistribution. Although capturing avalanches experimentally is challenging due to the competition between various timescales in active tissue, tracking the spatiotemporal evolution of local rearrangements is feasible within timescales where activities such as cell division are negligible. Such experiments could provide valuable evidence to support our analysis of rearrangement dynamics.  One such experiment is the fracturing of Trichoplax Adhaerens, during which regions with significant non-affine motion exhibit spatial correlations along the direction of the driving shear force  ~\cite{Prakash_2021}.}

The universality of tissue avalanches is not only reflected in the propagation of plastic events but also captured by a power-law distribution of avalanche sizes, with an exponent $\tau = -1.36$, strengthening the argument that epithelial tissues behave like plastic amorphous materials. We also propose a metric to not only predict tissue avalanches but also infer tissue stress in highly anisotropic systems based on an instantaneous static snapshot. In finite size system, the cut off avalanches size $S_c$ depends on the system size $N$ as $S_c \propto N^{d_f/d}$, where $d_f$ is {fractal dimension of avalanches} and $d$ is the dimension of the system  ~\cite{Lin_2014}. Future possible work can be performed with different system size to obtain the avalanches fractal dimension $d_f$  and further understand the finite size effect on avalanches.

We also propose a metric to not only infer tissue stress based on an instantaneous static snapshot but also predict tissue avalanches.
Quantification of tissue-level force and stress is necessary to understand the physics of many biological processes. However, direct measurement of stress in vivo is considerably challenging ~\cite{Haase2015,Bambardekar_Lenne2015,Bonnet_laser_ablation2012}. Compare to other non-invasive {methods} to estimate tissue stress such as Bayesian Force Inference  ~\cite{Ishihara_2012} and Variational Method for Image-Based Inference  ~\cite{Noll_Shraiman2020}, our approach using $C^*$ offers a simple and fast method to estimate tissue stress. {The advantage of $C^*$ lies in its independence from fitting parameters and system properties. Given a segmented movie of a monolayer undergoing deformation, $C^*$ can be easily derived from the segmented images and offers a decent estimate of tissue relative stress evolution during the process without requiring prior knowledge of the monolayer's mechanical properties.} {The downside of our approach is that it cannot provide a spatial distribution of local stress.} The strong agreement between $\sigma_{xy}$ and $C^*$ is noteworthy, especially since $C^*$ does not incorporate information about edge orientation, which directly affects stress. In an isotropic system, edge length alone is insufficient to infer stress. However, in a system undergoing large deformation and thus highly anisotropic, the influence of edge orientation diminishes, making edge length alone sufficient for stress inference. The impact of shape anisotropy is evident during the buildup phase or when the system is in a fluid state. In these scenarios, edges may have negligible tension, making edge tension independent of edge length. Consequently, it is possible that systems with similar $C^*$ values could exhibit significantly different stress levels. 

The impact of triggers on avalanches goes beyond simply initiating them; we observed a significant dependence of avalanche size on the trigger location. By manually shrinking vanishing edges at various locations within the same configuration, we noted that the size of the resulting avalanches varied markedly. This indicates that the location of the initial excitation has a profound influence on the final size of the avalanche. A promising future research direction could involve developing a theoretical framework that moves beyond the mean field approach to more accurately capture the spatial heterogeneity in the tissue.

\section*{Method} 
\subsection*{Simulation Model}

To numerically study the behavior of dense epithelial tissues under large deformation, we use a Voronoi-based version  ~\cite{Bi_PRX_2016} of the Vertex Model  ~\cite{Fletcher_BPJ_2014_vertex_review} , where the degrees of freedom are the set of cell centers denoted as $\{\bm r_i\}$ and the geometric configurations of cells are derived from their respective Voronoi tessellation. The  biomechanics governing interactions both within and between cells can be effectively represented at a coarse-grained level  ~\cite{Farhadifar_CB_2007, Staple_EPJE_2010}, expressed in terms of a mechanical energy functional associated with cell shapes, given by:
\begin{equation}
E= \sum_{i=1}^N \left[ K_A (A_i-A_0)^2+ K_P (P_i-P_0)^2 \right],
\end{equation}
where $A_i$ and $P_i$ represent the area and perimeter of the $i$-th cell, respectively. The parameters $K_A$ and $K_P$ denote the area and perimeter moduli, respectively. The values $A_{0}$ and $P_{0}$ correspond to the preferred area and perimeter values, with $A_{0}$ specifically set to the average area per cell $\bar{A}$. Without the loss of generality, we choose $K_P A_{0}$ as the energy unit and $\sqrt{A_{0}}$ as the length unit. This leads to the dimensionless form of the energy
\begin{equation}
\varepsilon = \sum_{i=1}^N \kappa_A (a_i - 1)^2 + (p_i - p_0)^2,
\label{energy_func}
\end{equation}
where $\kappa_A=K_A A_{0}/K_P$ represents the rescaled area elasticity, governing the cell area stiffness relative to the perimeter stiffness, {$a_i=A_i/\sqrt{A_0}$ and $p_i=P_i/\sqrt{A_0}$ are the rescaled area and perimeter, respectively,} and $p_0=P_0/\sqrt{A_{0}}$ the cell shape index. {In this particular work, we studied only the regime with $K_A=0$ and $K_P=1$.} To investigate the mechanical response of the tissue, we apply simple-shear deformation to the simulated tissues utilizing Lees-Edwards boundary conditions  ~\cite{Lees_Edwards_1972}. Initially, strain-free configurations ($\gamma = 0$) are generated with randomly distributed cell centers. The FIRE algorithm  ~\cite{fire_algo} is subsequently employed to minimize the energy functional in accordance with Eq. \ref{energy_func}. Strain is then incrementally applied in steps of $\Delta\gamma = 2 \times 10^{-3}$ until reaching a maximum value of $\gamma_{max} = 6$. Alongside the modification of periodic boundary conditions to account for the strain, an affine displacement field $\Delta \mathbf{r_i} = \Delta\gamma \ y_i \ \hat{x}$ is applied to the cell centers. Following each increment of strain, the FIRE algorithm is again utilized to minimize the energy functional (Eq.\ref{energy_func}) until the residual forces acting on cell centers fall below $10^{-14}$, so that all resultant tissue states are meta-stable. This procedural approach effectively corresponds to investigating the system within the athermal quasi-static limit ($\dot{\gamma} \to 0$). {The tissues under examination consist of cell populations with $N=400$, accompanied by varying cell shape indices $p_0$. A total of $84$ random initial samples were simulated for each set of parameter values. The shape indices were incrementally varied from $3.66$ to $4$, with a step size of $0.2$, except near the rigidity transition $p_0^* = 3.81$. For $3.8 \leq p_0 \leq 3.82$, the parameter was incremented with a finer step size of $0.02$ to study the transition.}
 We calculate the tension, denoted as $\mathbf{T}_{ij}$, acting along an edge $\mathbf{l}_{ij}$ shared by cells $i$ and $j$ using the equation  ~\cite{Yang_PNAS_2017,Ishihara_2012,Chiou_2012}
\begin{equation}
\mathbf{T}_{ij} = \frac{\partial \varepsilon}{\partial \mathbf{l}_{ij}}=2[(p_i-p_0) + (p_j-p_0)]  \hat{\mathbf{l}}_{ij},
\label{def:tension}
\end{equation}
where $\hat{\mathbf{l}}_{ij}$ represents the unit vector along $\mathbf{l}_{ij}$. Furthermore, the global tissue shear stress $\sigma$ can be obtained by 
\begin{equation}
\sigma = \sigma_{xy}\equiv \dfrac{1}{N} \sum_{i<j} T_{ij}^x   l_{ij}^y,
\label{def:stress}
\end{equation}
where $T_{ij}^{x}$ denotes the $x$-component of $\mathbf{T}_{ij}$ and $l_{ij}^{y}$ stands for the $y$-component of $\mathbf{l}_{ij}$. 

\section*{Data Availability}
The data that support the findings of this study are available from the corresponding
author upon reasonable request.

\section*{Code Availability}
The custom code developed for this study is available at GitHub Repository (\url{https://github.com/jxhuangphys/sheared_voronoi_model}) under the MIT license. For further inquiries, please contact the corresponding author.

\section*{Acknowledgments}
We thank Craig E. Maloney, Marko Popović, and Jie Lin for insightful discussions. 
We acknowledge support from the National Science Foundation (grant nos. DMR-2046683 and PHY-2019745), the Alfred P. Sloan Foundation, the Human Frontier Science Program (Ref.-No.: RGP0007/2022), the NIGMS of the National Institutes of Health (NIH) under award number R35GM15049, and the Northeastern University Discovery Cluster.  

\section*{Author Contributions}
All authors contributed to the research design. A.Q.N. and J.H. performed and analyzed the numerical simulations with guidance from D.B. The manuscript was written by A.Q.N. and D.B., with input from J.H.

\renewcommand{\theequation}{S.\arabic{equation}}
\setcounter{equation}{0}
\begin{center}
    \fontsize{18}{15}\selectfont{Supplementary Text for} \\ 
    {\fontsize{14}{15}\selectfont{ \bf{Origin of yield stress and mechanical plasticity in model biological tissues}}}
    \\
\end{center}
\subsection*{Computing tissue level linear mechanical response}
The tissue level mechanical response was quantified by the shear modulus $G$. We computed $G$ using Born-Huang formulation in the limit of infinitesimal affine strain $\gamma$  ~\cite{Maloney_2006,li2019mechanical}: 
\begin{equation}
    G = \frac{1}{A_{total}}\big(\frac{\partial^2 E}{\partial \gamma ^2}-\Xi_{i\alpha}H^{-1}_{i\alpha j\beta}\Xi_{j\beta}\big),
    \label{shear modulus}
\end{equation}
where the Roman indexes $i,j$ label cells and Greek indexes $\alpha,\beta$ denote Cartesian components. $\Xi_{i\alpha}$ is the derivative of the force on cell $i$th with respect to the strain $\gamma$:
\begin{equation}
    \Xi_{i\alpha} = \frac{\partial^2 E}{\partial r_{i\alpha}\partial \gamma}
\end{equation}
$H$ is the Hessian matrix given by the second derivative of the tissue energy E with respect to position cells position:
\begin{equation}
    H_{i\alpha j\beta }=\frac{\partial^2 E}{\partial r_{i\alpha}\partial r_{j\beta }}
\end{equation}
\subsection*{Steady state solution and asymptotic behavior in dual-state SGR}
\label{SGR section}

In the Fokker-Planck equation of motion {(Eqn.\eqref{Sto_DE} in the Main Text)}, the total yielding rate $\Gamma(t)$ {can be obtained}  ~\cite{SGR_Sollich_1998}:
\begin{equation}
\Gamma(t) = \Gamma_0 \int dE \  dl \: P(E,l,t) \exp\left[-\frac{E-kl^2/2}{x}\right]. \label{total_yielding_rate}   
\end{equation}
Since we are interested in the long-term steady shear, we look for a steady state solution to Eqn.\ref{Sto_DE}. In steady state, Eqn.\eqref{Sto_DE} becomes an ODE with respect to $l$:
\begin{equation*}
  \frac{\partial P}{\partial l} +\frac{\Gamma_0}{\dot\gamma } \exp\left[-\frac{E-kl^2/2}{x}\right] \ P =\frac{\Gamma}{\dot\gamma } \rho(E)\delta(l)
\end{equation*}
The steady-state solution is  ~\cite{SGR_Sollich_1998}:
\begin{equation}
    P(E,l)= \frac{\Gamma}{\dot\gamma}\rho(E)\exp(-ze^{-E/x})    
    \label{ss_sol}
\end{equation}

Where $z(l)$ is:
\begin{equation}
        z(l)=\frac{\Gamma_0}{\dot{\gamma}}\int_0^l dl'e^{kl'^2/2x}
\end{equation}

In steady state, the total yielding rate $\Gamma$ is just a constant and can be found by normalizing $P(E,l)$, giving the steady state solution for $P(E,l)$ of the form:
\begin{equation*}
    P(E,l)=\frac{\rho(E)\exp(-ze^{-E/x})}{\int_0^\infty  dl\: G_\rho(z)}
\end{equation*}
Where $G_\rho$ is:
\begin{equation*}
    G_\rho(z) = \int_0^\infty dE \:\rho(E)\exp(-ze^{-E/x})
\end{equation*}
$G_\rho$ can be separated into two parts. The first part, denoted by $G_\delta$, comes from the contribution of the zero energy traps (the Dirac-delta function in $\rho(E)$). $G_\Gamma$ denotes the second part coming from the non-zero energy traps (the Gamma function in $\rho(E)$):
\begin{align*}
    G_{\delta}(z) &= \int dE\: \delta(E)\exp(-ze^{-E/x})=\exp(-z)\\
    G_{\Gamma}(z) & = \int dE\: \frac{E^{\kappa-1}e^{-E/x_0}}{\Gamma(\kappa)x_0^\kappa}\exp(-ze^{-E/x}) \\
    G_\rho(z) &= f_0 G_{\delta}+(1-f_0)G_{\Gamma} 
\end{align*}
The steady-state solution in the long-time limit can be studied more conveniently using the following auxiliary functions:
\begin{align*}
    I_{0\delta} &= \int dl \: G_\delta\\
    I_{1\delta} &= k_\delta \int dl \: lG_\delta \approx 0\\
    I_{0\Gamma} &= \int dl \: G_\Gamma \\
    I_{1\Gamma} &= k\int dl \: lG_\Gamma \\   
\end{align*}

Since our main focus is the rheological response of the system, reflected in the macroscopic stress, we compute the system stress by ensemble averaging the local stress  ~\cite{SGR_Sollich_1998}:
\begin{equation}
    \sigma = \big<kl\big>=\int \int dE dl \: klP(E,l) \label{yield_stress_integral}
\end{equation}

Using the auxiliary functions and the steady-state solution, the stress is therefore given by:
\begin{align}
        \sigma &= \frac{(1-f_0)I_{1\Gamma}}{f_0I_{0\delta}+(1-f_0)I_{0\Gamma}}
        \label{stress solution}
\end{align}

For an element with zero yielding energy, the strain of the element is of order $\frac{\dot{\gamma}}{\Gamma_0}$. Therefore, in the low strain-rate limit, the strain of the element with zero yield energy is typically small so $G_\delta$ can be approximated in this limit as:
\begin{equation*}
    G_\delta = \exp(-l\Gamma_0/\dot{\gamma})
\end{equation*}
In the case of $\kappa = 2$ and let $\chi=\frac{x}{x_0}<2$, using the substitution $u=e^{-E/x}$ and integration by part, $G_\Gamma(z)$ can be integrated as follow:
\begin{align*}
    G_\Gamma(z)&=\int_1^0 \chi^2\ln(u)u^{\chi-1}\exp(-zu)du\\
    &=\frac{\chi^2(\chi-1)!}{z^\chi}\sum_{n=1}^\infty \frac{(-z)^n}{nn!} 
\end{align*}
The series $\sum_{n=1}^\infty \frac{(-z)^n}{nn!} $ converges by alternating series test and therefore $G_\Gamma(z)$ scales as $z^{-\chi}$. We then obtain the following scaling relations:
\begin{align}
    I_{0\delta }&\approx \dot\gamma/\Gamma_0\\
    I_{0\Gamma}&\approx C(\chi)\dot\gamma^{\chi}
    \label{scaling I0}\\
    I_{1\Gamma} &\approx D(\chi)\dot\gamma^{\chi}
    \label{scaling I1}
\end{align}

Using, the steady state solution, the proportion of time that elements spends in the zero yielding energy traps (fluid state) $P$ can be expressed in terms of $f_0$:
\begin{equation}
    P = \frac{f_0I_{0\delta}}{f_0I_{0\delta}+(1-f_0)I_{0\Gamma}} \label{fraction_and_f0}
\end{equation}
Combining Eq. \eqref{stress solution} and Eq. \eqref{fraction_and_f0}, the yield stress can be calculated as:
\begin{align}
    \sigma_y(x) &= \frac{(1-f_0)I_{1\Gamma}}{f_0I_{0\delta}+(1-f_0)I_{0\Gamma}}=(1-P)\frac{I_{1\Gamma}}{I_{0\Gamma}}=(1-P)\sigma_0 
    \label{yield_stress_sol}
\end{align}
Where $\sigma_0$ is the yield stress that arises solely from elements in a solid state, which is finite based on the scaling relation \eqref{scaling I0} and \eqref{scaling I1}.

\subsection*{Modeling tissue stress as a function of edge length critical cumulative distribution}
\renewcommand\thefigure{S\arabic{figure}} 
\setcounter{figure}{0}
\begin{figure}[htpb]
\includegraphics[width=1\columnwidth]{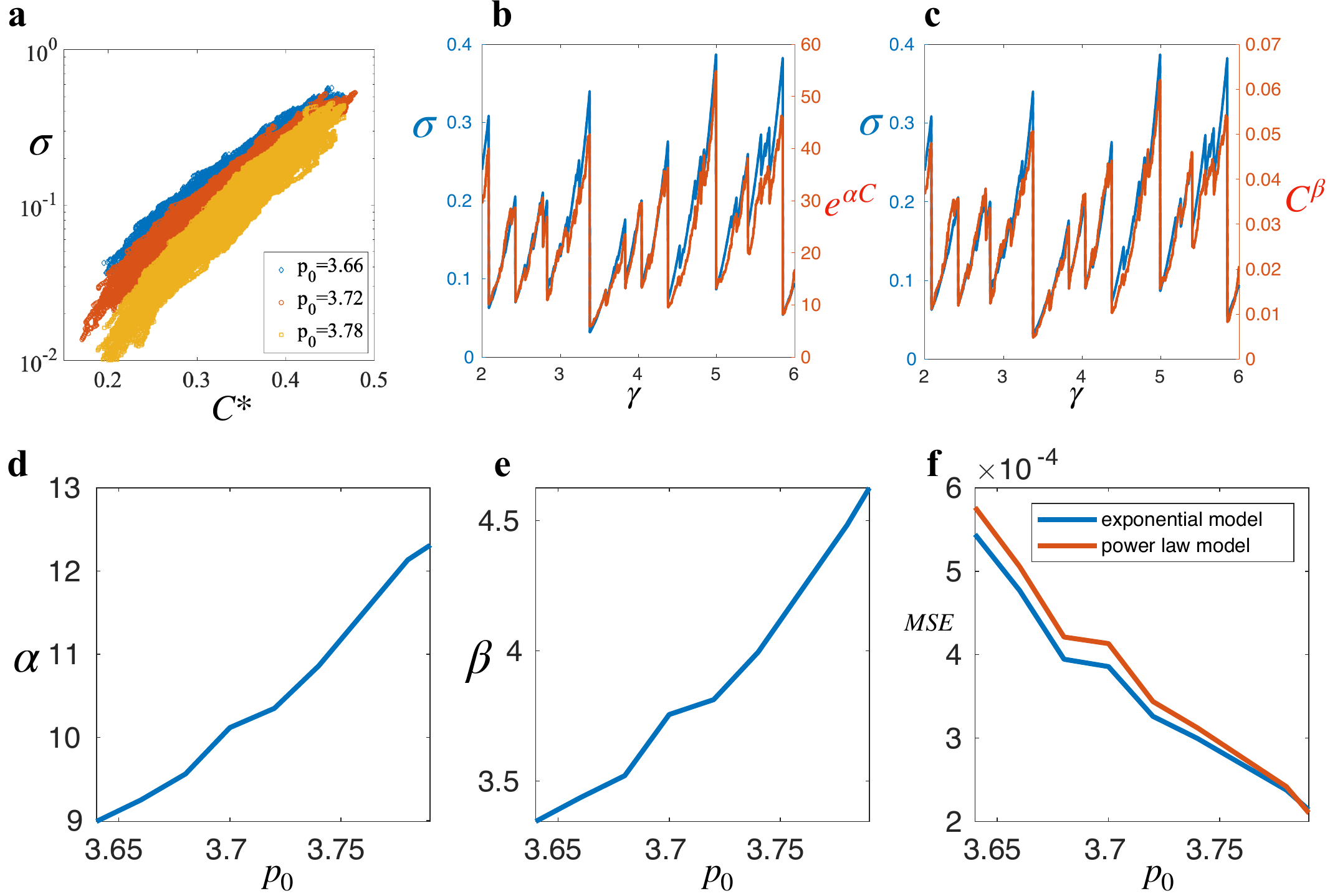}
\caption{Dependence of stress inference model fitting parameters $\alpha$ and $\beta$ on $p_0$.}

\label{fitting and error}
\end{figure}

{Plotting $C^*$ against $\sigma_{xy}$ for $p_0$ in the plastic regime reveals a clear exponential relationship (Fig.\ref{fitting and error}a). However, due to the limited range of $C^*$, distinguishing between an exponential and a power-law relationship remains challenging. To address this, we tested both the exponential model ($\sigma_{xy} = m_{exp} e^{\alpha C}$) and the power-law model ($\sigma_{xy} = m_{pow} C^\beta$). The only fitting parameters for each model are $\alpha$ and $\beta$, for the exponential and power-law models, respectively. As shown in figure \ref{fitting and error}(b,c), both models provide excellent estimates of tissue stress over a wide range of strain. The dependence of $\alpha$ and $\beta$ on $p_0$ is shown in figure \ref{fitting and error}(d,e), with both $\alpha$ and $\beta$ increasing as $p_0$ increases. To compare the two models, we first determined the unit conversion factors as $m_{exp} = \langle \overline{\sigma}/\overline{e^{\alpha C}} \rangle$ and $m_{pow} = \langle \overline{\sigma}/\overline{C^\beta} \rangle$. Here, $\langle ... \rangle$ represents ensemble averaging, and $\bar{...}$ represents averaging over different strains in a single simulation. The mean squared error (MSE) for both models was computed at different $p_0$ values, as shown in figure \ref{fitting and error}f. The MSE values indicate that the residuals are approximately two orders of magnitude smaller than the observed stress, demonstrating the high accuracy of both models. In addition, as a consequence of the decrease in yield stress with increasing $p_0$, the MSE also decreases as $p_0$ increases.}

\subsection*{An exponent of instability}
\begin{figure}[htpb]

\includegraphics[width=1\columnwidth]{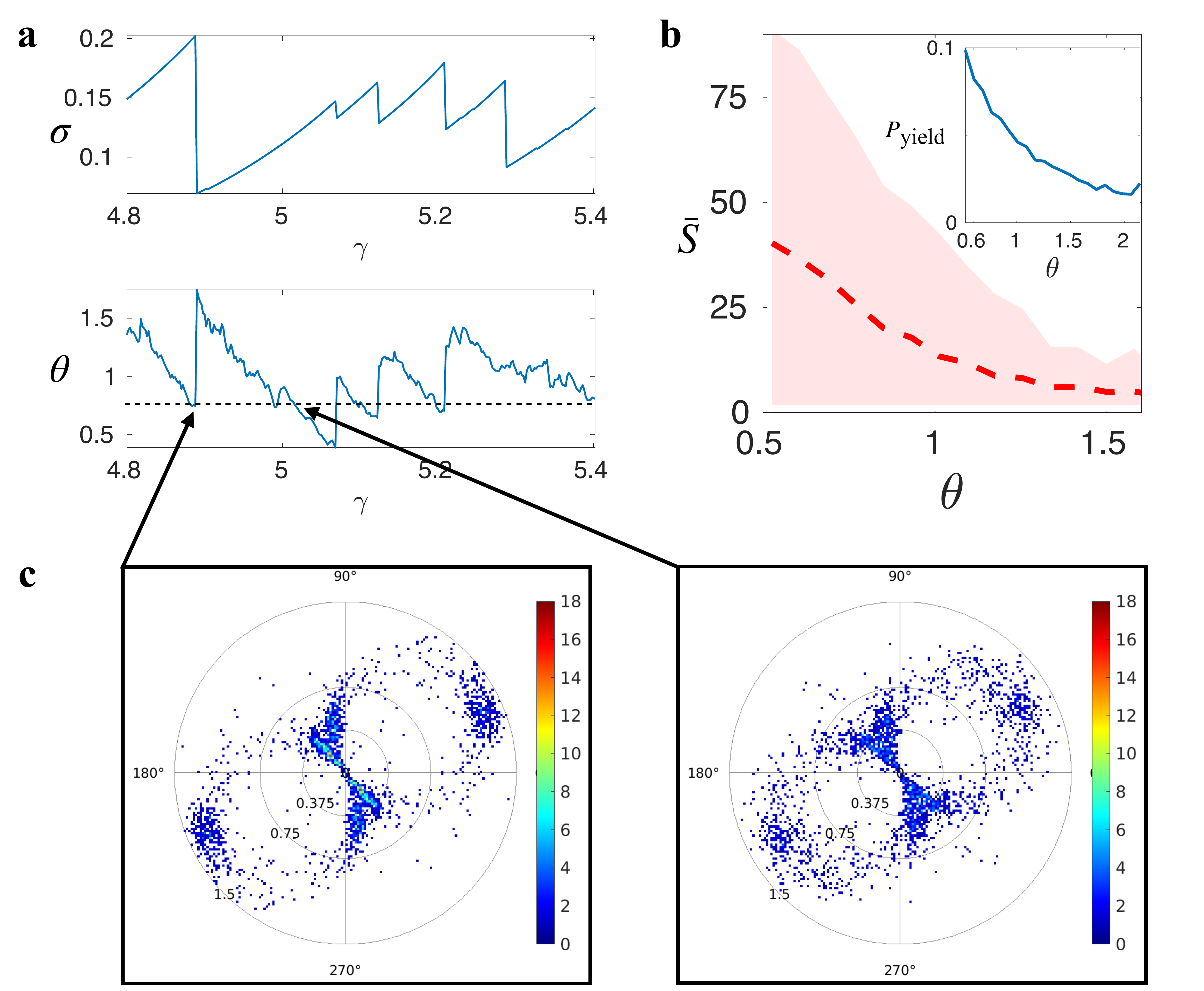}
\caption{A proposed exponent of instability.
(\textbf{a})An example of the evolution of tissue shear stress $\sigma_{xy}$ and $\theta$ exponent as strain increases. 
(\textbf{b})  Dependence of average avalanche size and the exponent of instability $\theta$. Inset: the probability to yield $\rho_+$ versus $\theta$. 
(\textbf{c}) The polar distribution of edges vector in an unstable and stable configuration. Data shown in Figure 3 is extracted from a system with shape index $p_0=3.74$ in the solid regime. }
\label{theta exponent}
\end{figure}
We extract the instability exponent, $\theta$, from the cumulative distribution function (c.d.f) of edge lengths, denoted as c.d.f($L$), by fitting a power law to c.d.f($L$) in the interval $0.05<L<0.5$. This interval specifically represents the short edges that are capable of undergoing T1 transitions.

We focused on tissues exclusively in the solid regime (shape index $p_0$ less than 3.81) and integrated this understanding with the dual-state coexistence proportion to extend the analysis to higher shape indices. For purely solid tissues, as stress builds up, the instability exponent $\theta$ gradually decreases. Conversely, when stress is relieved through avalanches, $\theta$ experiences a sharp increase (Fig.\ref{theta exponent}a), indicating that a significant number of soft spots are relaxed, making the system considerably more stable.

Furthermore, the $\theta$ exponent is correlated with avalanche properties, as evidenced by its relationship with the average avalanche size ($\bar{S}$) and the probability of avalanche occurrence. Systems with a lower $\theta$ exponent, indicating greater instability, tend to experience larger avalanches on average (Fig.\ref{theta exponent}b) and are more prone to yielding (Fig.\ref{theta exponent}b inset).

\subsection*{Normal modes unable to identify the soft spots in the Vertex-based model}
In amorphous solids, localization plays a crucial role in understanding the rheology of the material. An example of localization is shear transformation zone (STZ)  ~\cite{Falk_1998,Patinet_Falk_2016,LANGER_2006,Richard_2021,Manning_2007}, localized regions in which sudden and irreversible rearrangements occur when the material is subjected to shear. These STZ, also referred to as weak spots, can interact and lead to avalanches of irreversible plastic events, making the identification of these weak spots in disordered systems a crucial and challenging task. Research has indicated that the local yield stress could serve as a reliable predictor for these weak spots  ~\cite{Patinet_Falk_2016}. However, locally probing the system is impractical and does not align with our objective of making predictions based solely on current and historical snapshots. Another approach to this task involves analyzing the normal mode of the system near failure. Studies on systems of harmonic repulsive particles have demonstrated that low-frequency modes typically correspond to low energy barriers  ~\cite{Xu_Liu_epl2010}, making them dominant modes during a plastic event  ~\cite{Brito_Wyart_JSM2007,Maloney_2004,Xu_Liu_epl2010}. Furthermore, in systems with explicit separation dependence potential (such as Hertzian and Lennard-Jones potentials), under the quasistatic limit, the evolution of low-frequency modes follows a distinct pattern: as the system approaches failure, a gradual decrease in frequency towards zero is anticipated  ~\cite{Manning_Liu_2011,Maloney_2004}.

\begin{figure}[htpb]
\includegraphics[trim= 0 2.5cm 0cm 0cm, width=\columnwidth]{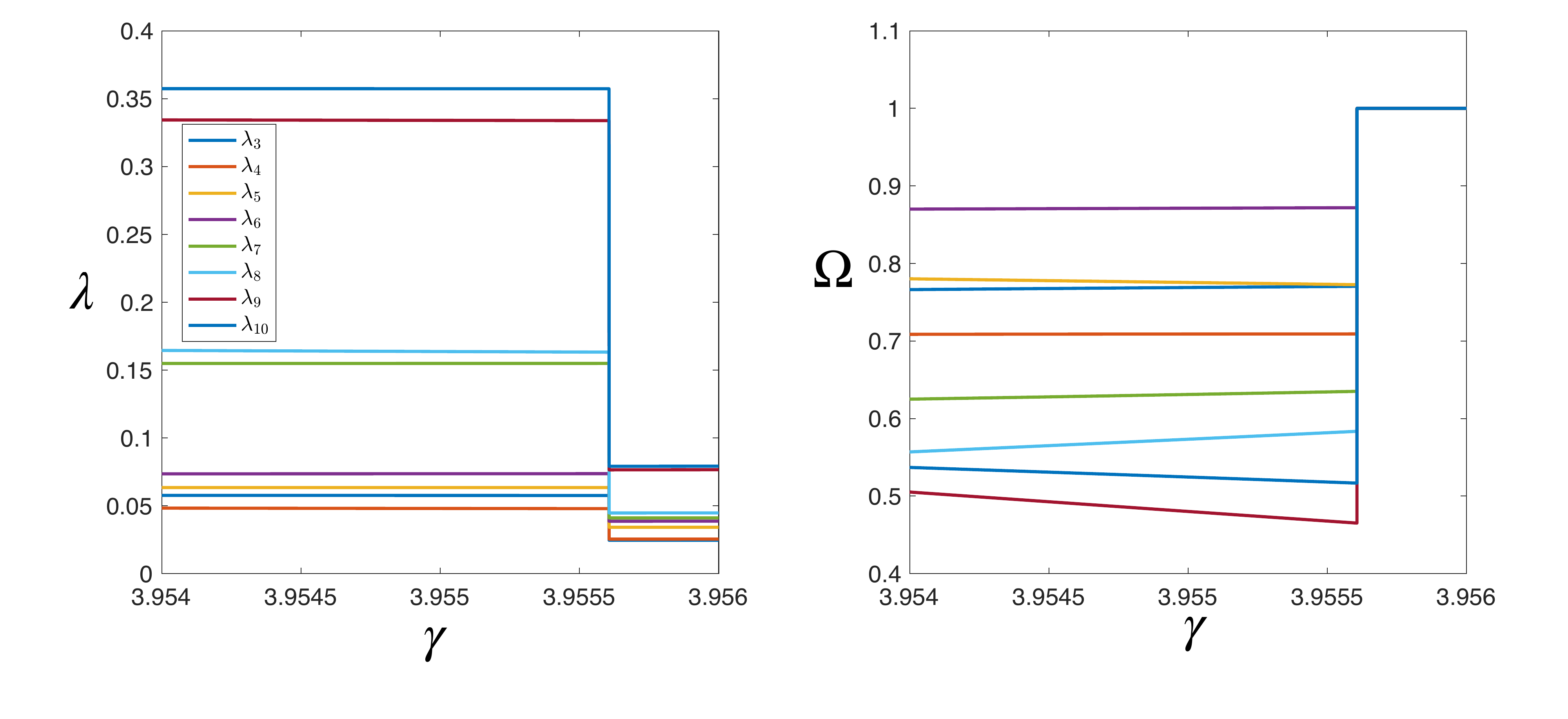}
\caption{Evolution of low-frequency mode near avalanches.}

\label{normal modes} 
\end{figure}

To see whether the normal modes of the Hessian could help to identify failure events, whenever there is a known avalanche in our simulation, we rerun the simulation starting at this particular strain but with the strain step decreased by 100 times and let the system approach to the avalanche again in the more detailed fashion. The low-frequency modes at the starting strain were then extracted from the Hessian. To keep track of the mode while the system evolves, we found the most similar mode to these starting low-frequency modes at each step and used them to represent the starting modes using the overlapping function $\Omega = \mathbf{e_i} . \mathbf{e_j} $, where $\mathbf{e_i}$ and $\mathbf{e_j}$ are eigenvectors of comparison. The overlap is shown on the right panel of Figure \ref{normal modes}. In contrast to what was observed in other systems, we did not see a gradual decrease in the low-frequency mode eigenvalues. Instead, it is always a sharp decrease but not zero right at the onset of avalanches, no matter how detailed we zoom in on the approaching process. This non-smooth sudden drop in the eigenvalues at the onset of avalanches arises from the cuspiness of the energy landscape. Because of the cuspiness, there is no saddle point when the system approaches a rearrangement event and therefore the curvature is always positive.

\begin{figure}[htpb]
\includegraphics[width=0.5\columnwidth]{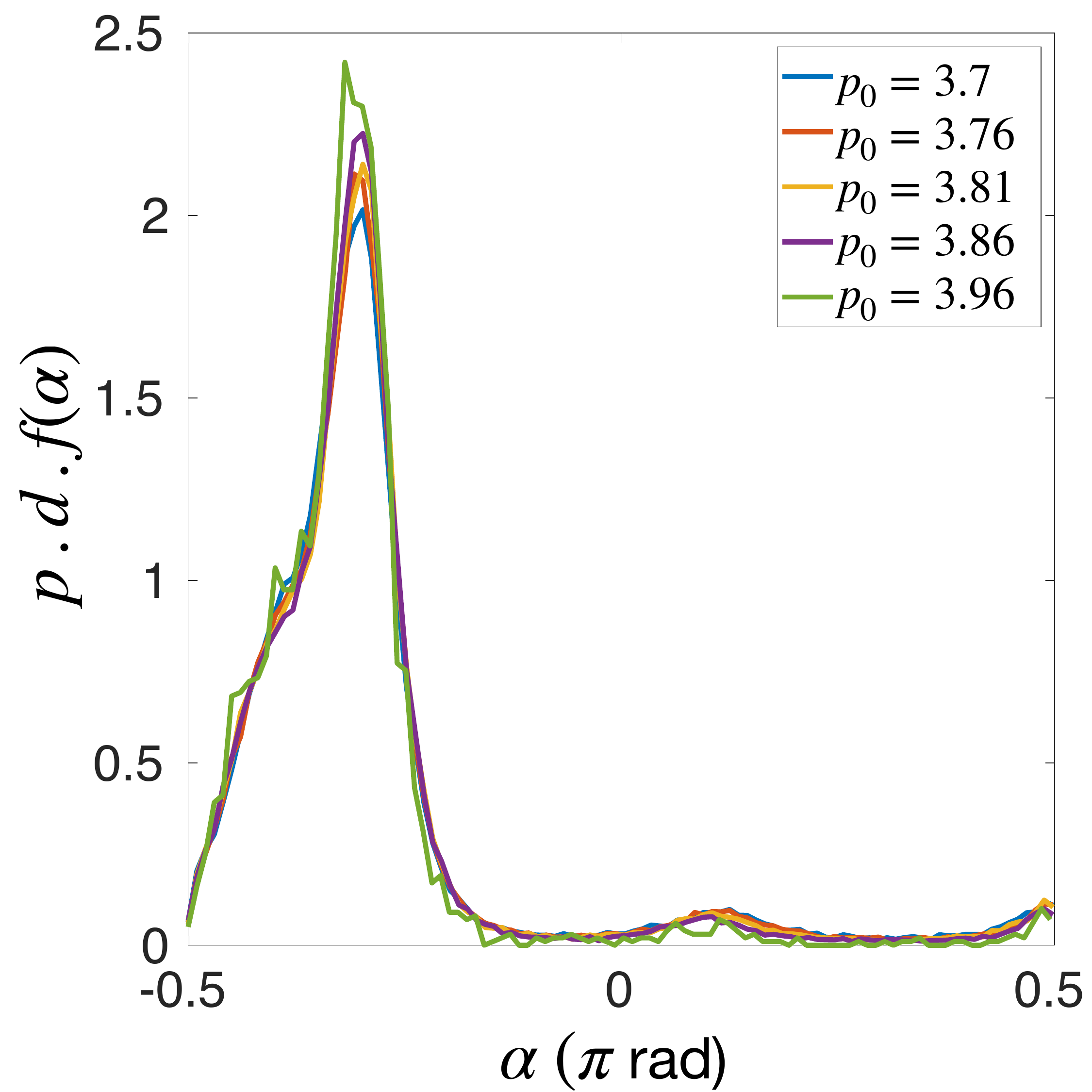}
\caption{The orientation distribution of T1-edges (edges undergo T1 transitions) at various values of $p_0$.}
\label{T1 orient}
\end{figure}

\bibliography{SGR.bib}

\end{document}